\documentstyle[aps,twocolumn]{revtex}

\global\parskip 6pt
\newcommand{\be}{\begin{equation}}
\newcommand{\ee}{\end{equation}}
\newcommand{\bea}{\begin{eqnarray}}
\newcommand{\eea}{\end{eqnarray}}
\newcommand{\non}{\nonumber}
\begin{document}
\twocolumn[\hsize\textwidth\columnwidth\hsize\csname
@twocolumnfalse\endcsname
\title{Conformal Field Theory Interpretation of Black Hole
Quasi-normal Modes}

\author{
Danny Birmingham\,${}^*$, Ivo Sachs\,${}^\dagger$ and
Sergey N. Solodukhin\,${}^\dagger$ }
\address{${}^*${\it
Department of Mathematical Physics,
University College Dublin,
Belfield, Dublin 4, Ireland.
}\\ ${}^\dagger${\it
Theoretische Physik,
Ludwig-Maximilians Universit\"{a}t,
Theresienstrasse 37, D-80333, M\"{u}nchen, Germany.}}
\maketitle

\begin{abstract}\noindent
  We obtain exact expressions for the quasi-normal modes of
various spin for the BTZ black hole. These modes determine the
relaxation time
of black hole perturbations. Exact agreement is found between
the quasi-normal frequencies and the location of the poles of the
retarded correlation function of the corresponding perturbations in the
dual conformal field theory. This then provides a new quantitative
test of the AdS/CFT correspondence.
    \\

{PACS: 04.70Dy, 04.60.Kz, 11.25.Hf\ \ \

}
\end{abstract}
\vskip2.pc]

The problem of how a perturbed thermodynamical system returns to equilibrium
is an important issue in statistical mechanics and finite temperature field
theory \cite{Kubo}.
For a small perturbation, this process is described by linear
response theory \cite{Kubo,Fetter}. The relaxation process is then
completely determined by the poles, in the momentum representation,
of the retarded correlation function of the perturbation. On the other hand,
black holes also constitute a thermodynamical system. At equilibrium, the
various thermodynamical quantities, such as the temperature and the entropy,
are determined in terms of the mass, charge and angular momentum of the black
hole. The decay of small perturbations of a black hole at equilibrium are
described by the so-called quasi-normal modes \cite{Frolov:wf}.
For asymptotically flat black hole space-times, quasi-normal modes
are analysed by solving the wave equation for matter or gravitational
perturbations, subject to the conditions that the flux at the horizon
is in-going, with
out-going flux at asymptotic infinity. The wave equation, subject
to these boundary conditions, admits only a discrete set of solutions
with complex frequencies. The imaginary part of these quasi-normal
frequencies then determine the decay time of small perturbations,
or equivalently, the relaxation of the system back to thermal equilibrium.

On another front, over the last few years increasing evidence has
accumulated that there is a correspondence between gravity and
quantum field theory in flat space-time
(for a review, see \cite{Aharony:1999ti}). In particular, this duality
has led to important progress in our understanding
of the microscopic physics of a class of near-extremal black holes.
The purpose of this letter is to analyse whether such a correspondence
exists between quasi-normal modes in anti-de Sitter (AdS)
black holes and linear
response theory in scale invariant finite temperature field theory.
A correspondence between quasi-normal modes and the decay of
perturbations in the dual conformal field theory (CFT) was first suggested
in \cite{Horowitz:2000jd}. The analysis of \cite{Horowitz:2000jd} is
based on the numerical computation of quasi-normal modes for
AdS-Schwarzschild black holes in four, five, and seven dimensions.
Further numerical computations of quasi-normal modes in asymptotically AdS
space-times have been presented in \cite{Robb}-\cite{Robb3}. For related
discussions in the context of black hole formation see
\cite{Danielsson:1999zt}. Qualitative agreement was found with the results expected
from the conformal field theory side.
However, a quantitative test of such a correspondence
between quasi-normal modes and the linear response of the dual conformal
field theory is lacking so far.
In this paper, we consider the $2+1$ dimensional AdS
black hole \cite{BTZ}, and
show that there is a precise quantitative agreement
between its quasi-normal
frequencies and the location of the poles of the
retarded correlation function describing the linear response on
the conformal field theory side. Both computations are performed
analytically. As a result, we can identify not just the lowest,
but the complete (infinite) set of frequencies on both sides of
the AdS/CFT correspondence. In spite of
its simplicity, this model plays an
important role also for black holes in
higher dimensions whose near-horizon geometry is AdS$_{2+1}$
(see e.g. \cite{Birmingham:2001dt} for a review and references).

The metric of the BTZ black hole is given by
\begin{eqnarray}
ds^2&=&-\sinh^2\mu\left(r_+dt-r_-d\phi\right)^2+
d\mu^2\nonumber\\
&&+\cosh^2\mu\left(-r_-dt+r_+d\phi\right)^2\ .
\label{4}
\end{eqnarray}
The angular coordinate $\phi$ has period $2\pi$, and the radii of the
inner and outer horizons are denoted by $r_+$ and $r_-$, respectively.
We have also set to unity the radius of the anti-de Sitter space,
$\ell\! \equiv \!1$.
The dual conformal field theory on the boundary is $1+1$ dimensional,
the conformal symmetry being generated by two copies of
the Virasoro algebra acting separately on left- and right-moving sectors.
Consequently,  the conformal field theory splits into two
independent sectors at thermal equilibrium with temperatures
\begin{equation}
T_L=(r_+-r_-)/2\pi\ , \quad T_R=(r_++r_-)/2\pi\ .
\label{5}
\end{equation}
According to the AdS${}_3$/CFT${}_2$ correspondence,
to each field of spin $s$  propagating in AdS${}_3$ there
corresponds an operator $\cal O$ in the dual conformal field theory
characterised by conformal weights $(h_L,h_R)$ with \cite{Aharony:1999ti}
\begin{equation}
h_R+h_L=\Delta \ ,\quad h_R-h_L=\pm s\ ,
\label{1}
\end{equation}
and $\Delta $ is determined in terms of the mass $m$ of the field.
In particular, we have
\begin{equation}
\Delta=1+\sqrt{1+m^2}\ ,
\label{2}
\end{equation}
for scalar fields, and
\begin{equation}
\Delta=1+|m|\ ,
\label{3}
\end{equation}
for both fermionic and vector fields. For a small perturbation,
the manner in which the field theory relaxes back to thermal equilibrium
can then be analysed within linear response
theory \cite{Fetter}. One expects that at late times the
perturbed system will approach equilibrium exponentially with a
characteristic time-scale. This time-scale is inversely proportional
to the imaginary part of the poles, in momentum space,
of the correlation function of the perturbation operator
${\cal O}$. In the present case, according
to
our proposal, the relevant correlation function is the retarded real time
correlation function
\begin{eqnarray}
\label{GR11}
{\cal D}^{\mathrm{ret}}(x,x')&=&i\theta(t-t')\left< [{\cal O}(x),
{\cal O} (x')]\right>_{T}\non\\
&=&i\theta(t-t^{\prime})\bar{\cal D}(x, x^{\prime})\ ,
\label{6}
\end{eqnarray}
where $\bar{\cal D}(x,x^{\prime}) = {\cal D}_{+}(x, x^{\prime})
-{\cal D}_{-}(x, x^{\prime})$ is the commutator evaluated in
the equilibrium canonical ensemble. For a conformal
field theory at zero temperature, the 2-point correlation functions can
be determined,
up to a normalisation, from conformal invariance.
At finite temperature $T$, one has to take into account
the infinite sum over images to render
the correlation function periodic in imaginary time, with period $1/T$.
The result of this summation in two dimensions has
been determined in \cite{Cardy}, and depends only on the conformal
dimensions $(h_L,h_R)$ of the perturbation operator.
We have ($x^\pm\!=\!t\pm \sigma$),
\begin{eqnarray}
{\cal D}_+(x)=\frac{(\pi T_R)^{2h_R}}{\sinh^{2h_R}
(\pi T_R x^- \!-\!i\epsilon)}
\frac{(\pi T_L)^{2h_L}}{\sinh^{2h_L}(\pi T_L x^+\! -\!i\epsilon)}
\label{7}\nonumber
\end{eqnarray}
and a similar expression for ${\cal D}_-(x)$ with $\epsilon \to -\epsilon$.
In order to determine the location of the poles, we need to compute the
Fourier transform of  (\ref{GR11}). This is complicated
by the presence of the $\theta$ function. We can, however determine the
location of the poles indirectly. For this we first consider the
Fourier transform of the commutator $\bar{\cal D}(x)$.
This integral can be evaluated using contour techniques, leading
to \cite{Gubser}
\begin{eqnarray}
&&\bar{\cal D}(k_+,k_-) \propto
\Gamma\left(h_L+i\frac{p_+}{2\pi T_L}\right)
\Gamma\left(h_R+i\frac{p_-}{2\pi T_R}\right)\times\nonumber\\
&&\qquad\qquad\Gamma\left(h_L-i\frac{p_+}{2\pi T_L}\right)
\Gamma\left(h_R-i\frac{p_-}{2\pi T_R}\right)\ ,
\label{9}
\end{eqnarray}
where $p_\pm\!=\!\frac{1}{2}(\omega\mp k)$. This function has poles
in both the upper and lower half of the $\omega$-plane.
The poles lying in the lower half-plane are the same as the poles of the
retarded correlation function (\ref{6}).
Restricting the poles of (\ref{9}) to the lower half-plane,
we find two sets of poles
\begin{eqnarray}
\omega_L&=&k-4\pi i T_L(n+h_L)  \ ,\nonumber\\
\omega_R&=&-k-4\pi i T_R(n+h_R)\ .
\label{10}
\end{eqnarray}
Here, and in the following, $n$ takes the integer values $(n=0,1,2,...)$.
This  set of poles characterises the decay
of the perturbation on the CFT side,
and coincides precisely with the quasi-normal frequencies
of the BTZ black hole, as we shall now show for fields of various spin.

\noindent \underline{\it Scalar Perturbation} ($s=0$). Scalar
perturbations are
described by the wave equation
\begin{equation}
\left(\nabla^2- m^2\right)\Phi=0\ .
\label{s1}
\end{equation}
We use the ansatz
\begin{equation}
\Phi=e^{-i(k_+x^++k_-x^-)}R(\mu)\ ,
\label{s2}
\end{equation}
where $x^+=r_+t-r_-\phi$, $ x^-=r_+\phi-r_-t$, and
\begin{eqnarray}
(k_++k_-)(r_+-r_-)&=&\omega-k \ , \nonumber \\
(k_+-k_-)(r_++r_-)&=&\omega+k \ .
\label{s3}
\end{eqnarray}
Here, $\omega$ and $k$ are the energy and angular momentum of the
perturbation. Changing variables to $z=\tanh^2\mu $,
we end up with the hypergeometric equation
\begin{eqnarray}
z(1-z)\frac{\hbox{d}^2R}{\hbox{d} z^2}&+&(1-z)\frac{\hbox{d}
R}{\hbox{d} z}\nonumber\\
&+&\left[\frac{k_+^2}{4z}
-\frac{k_-^2}{4}-\frac{m^2}{4(1-z)}\right]R=0\ .
\label{s5}
\end{eqnarray}
The solution which is in-going at the horizon is given by
\begin{equation}
R(z)=z^\alpha(1-z)^{\beta_s} F(a_s,b_s,c_s,z)\ ,
\label{s6}
\end{equation}
where $\alpha=-\frac{ik_+}{2}$,
$\beta_s=\frac{1}{2}\left(1-\sqrt{1+m^2}\right)$, and
\begin{eqnarray}
a_s&=&\frac{(k_+-k_-)}{2i}+\beta_s\ ,\quad
b_s=\frac{(k_++k_-)}{2i}+\beta_s\ ,\non\\
 c_s&=&1+2\alpha\ .
\label{s7}
\end{eqnarray}
The quasi-normal modes for the scalar
perturbations were found in \cite{Danny}
by imposing the vanishing Dirichlet condition at infinity.
Here, we re-evaluate these modes using the condition that
the flux given by
\begin{eqnarray}
{\cal F}=\sqrt{g}\frac{1}{2i}\left(R^*\partial_\mu R
- R\partial_{\mu}R^{*}\right)
\label{s8}
\end{eqnarray}
vanishes at asymptotic infinity.
For $m^2>0$, the asymptotic flux has
a set of divergent terms, with the leading term of order $(1-z)^{2\beta_s}$.
Each of these terms is proportional to \cite{AS}
\begin{equation}
\left|\frac{\Gamma(c_s)
\Gamma(c_s-a_s -b_s)}{\Gamma(c_s-a_s )\Gamma(c_s-b_s )}\right|^2\ .
\label{s10}
\end{equation}
Thus, the asymptotic flux vanishes if $c_s-a_s=-n$, or $c_s-b_s=-n $, i.e.
\begin{eqnarray}
\frac{i}{2}(k_+\pm k_-)&=&n+\frac{1}{2}
\left(1+\sqrt{1+m^2}\right)\ .
\label{s12}
\end{eqnarray}
Using (\ref{s3}), one sees that these are the quasi-normal
modes found in \cite{Danny}.
For the scalar bulk field, we have
$h_L=h_R=\frac{1}{2}\left(1+\sqrt{1+m^2}\right)$.
Thus, we observe that (\ref{s12}) exactly reproduces (\ref{10}).

In AdS space-time, a negative mass squared for a scalar
field is consistent,  as long as $-1<m^2<0$. A detailed analysis shows that
in this case there is a second set of modes with
$a_s=-n$, or $ b_s=-n $, that is
$h_L=h_R=\frac{1}{2}\left(1-\sqrt{1+m^2}\right)$. This is in fact
expected as for $-1<m^2<0$ there
are two sets of dual operators with $\Delta_+=1+\sqrt{1+m^2}$ and
$\Delta_-=1-\sqrt{1+m^2}$, respectively \cite{Aharony:1999ti}.
The second set of quasi-normal frequencies in this range then matches exactly
the dual operators with $\Delta=\Delta_-$.
We note in passing that the Dirichlet boundary condition
suggested in \cite{Horowitz:2000jd} leads to the same quasi-normal
modes for $m^2>0$ but does not lead to any quasi-normal modes for $m^2<0$.

\noindent \underline{{\it Fermion Perturbation}} $(s=1/2)$.
In \cite{Cardoso}, the quasi-normal fermionic perturbation in the BTZ
background has been analysed numerically.
However, as expected, it is possible
to find analytic solutions in this simple
case \cite{Das}. We begin with the
Dirac equation
\begin{equation}
(D\!\!\!\!/+m)\Psi=0\ ,\,\,
\Psi=e^{-i(k_+x^++k_-x^-)}\pmatrix{\psi_1\cr\psi_2}\ .
\end{equation}
Following \cite{Das}, we make the substitutions
\begin{eqnarray}
\psi_1\pm\psi_2&=&\sqrt{\frac{\cosh \mu \pm\sinh\mu}{\cosh \mu \sinh\mu}}
(\psi_1'\pm\psi_2')\ ,
\label{f1}
\end{eqnarray}
to obtain
\begin{eqnarray}
&&2(1-z)z^{1/2}\partial_z\psi_1'
+i(k_+ z^{-1/2}+k_- z^{1/2})\psi_1'=\nonumber\\
&&\qquad\qquad\qquad\quad-\left[i(k_++k_-)+m+{1\over 2}\right]\psi_2'\ ,
\label{f4}
\end{eqnarray}
and a similar equation where
$\psi_1'$, $\psi_2'$, and $k_\pm$ and $-k_\pm$ are interchanged.
The solutions of these equations
with in-going flux at the horizon are given by
\begin{eqnarray}\label{p1}
\psi_1'&=&z^\alpha (1-z)^{\beta_f} F(a_f,b_f,c_f,z)\ ,\\
\psi_2'&=&({a_f-c_f\over c_f}) z^{\alpha + 1/2}
(1-z)^{\beta_f} F(a_f,b_f+1,c_f+1,z),\nonumber
\end{eqnarray}
where $\alpha=-{ik_+\over 2}$,
$\beta_f=- {1\over 2}(m+{1\over 2})$, $c_f={1\over 2}+2\alpha$, and
\begin{eqnarray}
&&a_f=\frac{k_{+} - k_{-}}{2i}+ \beta_f + \frac{1}{2}\ ,
\quad b_f=\frac{k_{+} + k_{-}}{2i} + \beta_f\ .
\label{f6}
\end{eqnarray}
The asymptotic
form ($z\to 1$) of $\psi_1$ and $\psi_2$ can now be determined explicitly.
In analogy with the scalar perturbations, we then impose the condition
that the flux \cite{Das}
\begin{equation}
{\cal F}=\sqrt{g}
\bar\Psi\gamma_\mu\Psi\simeq(1-z)^{-1}(|\psi_1|^2-|\psi_2|^2)
\end{equation}
vanishes at infinity.
The resulting quasi-normal modes are then obtained
as follows. For $m>0$, the
leading divergent term in the asymptotic flux is of order
$(1-z)^{2\beta+ 1}$. Vanishing flux then requires that the
coefficient
\begin{equation}
{\Gamma (c_f)\Gamma (c_f-a_f-b_f)\over \Gamma
(c_f-a_f)\Gamma (c_f-b_f)}\ ,
\end{equation}
vanishes, that is,
\begin{eqnarray}
\frac{i}{2}(k_++k_-)&=&n+\frac{1}{4}+\frac{m}{2}\quad\hbox{or}\nonumber\\
\frac{i}{2}(k_+-k_-)&=&n+\frac{3}{4}+\frac{m}{2}\ .
\end{eqnarray}
The above conditions imply that all coefficients of the
sub-leading, asymptotically non-vanishing, contributions to the flux also
vanish. Thus, we have precise agreement with (\ref{10}),
where the left and right conformal weights are given by
$h_{L} = \frac{1}{4} + \frac{1}{2}m$, and $h_{R} = \frac{3}{4}
+ \frac{1}{2}m$. For $m<0$, one obtains a similar result with
$h_{L} = \frac{3}{4} - \frac{1}{2}m$,
and $h_{R} = \frac{1}{4} - \frac{1}{2}m$.
Note again that imposing Dirichlet boundary conditions
for $\psi_1$ and $\psi_2$ at infinity would lead to
the absence of quasi-normal modes for $-1<m<1$. We can
think of no physical reason for the absence of quasi-normal modes in this
range of masses. Thus, we take this as another motivation for imposing
vanishing flux at infinity, rather that Dirichlet conditions
for asymptotically AdS space-times. One should also note that
for positive mass the spinor perturbation is asymptotically
left-handed, whereas it is right-handed for negative mass (see also
\cite{Aharony:1999ti}).

\noindent \underline{{\it Vector Perturbation}} $(s=1)$.
The massive Maxwell field in AdS${}_3$ is described by the
first order equation
\begin{equation}
\epsilon_\lambda^{\;\;\alpha\beta}\partial_\alpha A_\beta =-mA_\lambda\ .
\label{v2}
\end{equation}
Let
\begin{equation}
A_i=e^{-i(k_+x^++k_-x^-)}A_i(\mu)\ ,
\label{v3}
\end{equation}
where $A_{1,2}=A_+\pm A_-$. Then, after changing variables as before,
 we recover the scalar equation (\ref{s5}) for $A_1$ and $A_2$
\cite{Das}, where the scalar mass squared is replaced by
$m^2+2\varepsilon_i m$, with $\varepsilon_1=-\varepsilon_2=1$.
As in the scalar case, the solutions with in-going
flux at the horizon are then given by
\begin{eqnarray}
A_1&=&e_1\;z^\alpha(1-z)^{\beta_v+1}F(a_v+1,b_v+1,c_v,z)\ ,\quad\nonumber\\
A_2&=&e_2\;z^\alpha(1-z)^{\beta_v}F(a_v,b_v,c_v,z)\ ,
\label{v5}
\end{eqnarray}
where $\alpha=-{ik_+\over 2}$, $\beta_v=\frac{m}{2}$, $c_v=1+2\alpha$, and
\begin{eqnarray}
a_v&=&\frac{(k_+-k_-)}{2i}+\beta_v\ ,\quad
b_v=\frac{(k_++k_-)}{2i}+\beta_v\ .
\label{v6}
\end{eqnarray}
Note that the two \lq\lq scalar modes" in (\ref{v3})
are not independent. The
first order equation (\ref{v2}) relates the two coefficients $e_1$ and
$e_2$ by
\begin{equation}
\frac{e_2}{e_1}=\frac{i(k_+-k_-)+m}{i(k_++k_-)-m}
= \frac{c_v-b_v-1}{b_v}\ .
\label{v7}
\end{equation}
The remaining component $A_\mu$ is related to $A_\pm$ by
\begin{equation}
A_\mu=\frac{1}{m
\cosh\mu\sinh\mu}\partial_{[+}A_{-]}\ .
\label{v8}
\end{equation}
For a real vector field, the particle flux is not defined.
One way to avoid this difficulty is to consider a complex vector field.
Alternatively, one can consider the energy flux divided by the red-shifted
frequency \cite{Gubser}. Both approaches lead to the same conditions,
namely that $A_{1}$ and $A_2$ vanish at infinity. Thus, we impose
Dirichlet boundary condition for $A_{1,2}$.
Using (\ref{v7}), one finds the leading asymptotic behaviour
($z \rightarrow 1$) of the
solutions (\ref{v5}) for positive $m$ is given by
\begin{eqnarray} \label{v9}
A_1&\simeq&  (1-z)^{-{m\over 2}} \frac{\Gamma(c_v) \Gamma(a_v+b_v-c_v+2)}
{\Gamma(a_v +1)\Gamma(b_v)}\ ,\\
A_2&\simeq& (1-z)^{1-{m\over 2}}(c_v-b_v-1) \frac{\Gamma(c_v)
\Gamma(a_v +b_v -c_v)}{\Gamma(a_v)\Gamma(b_v)}\ . \nonumber
\end{eqnarray}
By imposing the vanishing Dirichlet condition at infinity for the
components $A_1$ and $A_2$, we find the quasi-normal modes $a_v+1=-n$, or
$b_v=-n$, i.e.
\begin{eqnarray}
\frac{i}{2}(k_+-k_-)&=&n+1+\frac{m}{2}\quad\hbox{or}\nonumber\\
\frac{i}{2}(k_++k_-)&=&n+\frac{m}{2}\ .
\label{v10}
\end{eqnarray}
Now, for spin $s=1$, the conformal weights are either
$|m|/2$ or $1+|m|/2$. Thus, we again find agreement with (\ref{10}), where the left and right conformal weights are given by
$h_{L} = \frac{1}{2}m$, and $h_{R} = 1+ \frac{1}{2}m$.
For negative $m$, the situation is analogous to the fermionic perturbations.
One finds the same conditions with $h_{L} = 1 - \frac{1}{2}m$,
and $h_{R} = - \frac{1}{2}m$.

In conclusion, we have shown that there is a quantitative agreement between
the quasi-normal frequencies of the BTZ black hole and the poles
of the retarded correlation function of
the corresponding perturbations of the
dual conformal field theory.
The relaxation time for the decay of the black
hole perturbation is determined by the
imaginary part of the lowest quasi-normal mode.
Our analysis thus establishes a direct relation between this relaxation
time and the time-scale for return to equilibrium of the dual conformal
field theory.
This result also provides a new quantitative
test of the AdS/CFT correspondence.

We would like to thank T.S. Evans, R. Myers,  A. Peet and H. Wagner
for helpful discussions.
D.B. would like to thank the
CERN Theory Division for hospitality during the initial stages of this work.
S.S. would like to thank the Aspen Center for Physics for hospitality
while this work was in progress. The work of D.B. and I.S. was partially
supported by Enterprise Ireland grant IC/2001/004. S.S. is supported by
the grant DFG-SPP 1096, Stringtheorie.

\end{document}